\documentstyle[11pt,aaspp4,epsfig]{article}

\begin{document}
\title{Temporal and Spectral Variabilities of 
High Energy Emission from Blazars
Using Synchrotron Self-Compton Models}

\author{Hui Li\altaffilmark{1} and
Masaaki Kusunose\altaffilmark{2}}  

\altaffiltext{1}{Theoretical Astrophysics (T-6, MS B288), 
Los Alamos National Laboratory, Los Alamos, NM 87545;
hli@lanl.gov}

\altaffiltext{2}{Department of Physics, School of Science,
Kwansei Gakuin University, Nishinomiya 662-8501, Japan; 
kusunose@kwansei.ac.jp}

\begin{abstract}
Multiwavelength observations of blazars such as Mrk 421 and 
Mrk 501 show that they exhibit strong short time variabilities
in flare-like phenomena. Based on the homogeneous 
synchrotron self-Compton (SSC)
model and assuming that time variability of the emission is
initiated by changes in the injection of nonthermal electrons,  
we perform detailed temporal and spectral studies of a purely
cooling plasma system, using 
parameters appropriate to blazars. 
One important parameter is the total injected energy ${\cal E}$ and 
we show how the synchrotron and Compton components respond 
as ${\cal E}$ varies.
When the synchrotron and SSC components have comparable peak fluxes,
we find that the SSC process contributes strongly to the electron
cooling and the whole system is nonlinear, thus simultaneously
solving electron and photon kinetic equations is necessary.
In the limit of the injection-dominated situation when the cooling
timescale is long, 
we find a unique set of model parameters that are fully constrained 
by observable quantities. 
In the limit of cooling-dominated situation, 
TeV emissions arise mostly from a cooled electron distribution and 
Compton scattering process is always in the Klein-Nishina regime,
which makes the TeV spectrum having a large curvature.
Furthermore, even in a single injection event, the multiwavelength
light-curves do not necessarily track each other because the electrons
that are responsible for those emissions might have quite different
lifetimes. 
We discuss in detail how one could infer important physical
parameters using the observed spectra. 
In particular, we could
infer the size of the emission region by looking for exponential
decay in the light curves.  We could also test the basic assumption
of SSC by measuring the difference in the rate of peak energy changes
of synchrotron and SSC peaks.
We also show that the trajectory in the photon-index--flux plane
evolves clockwise or counter-clockwise depending on the value of
${\cal E}$ and observed energy bands.

\end{abstract}


\section{INTRODUCTION}

Blazars are a class of flat radio spectrum, core-dominated active
galactic nuclei (AGNs).
The overall radiation spectra of blazars show two broad
peaks in the $\nu F_{\nu}$ space; one is between IR and X-rays,
and the other in the $\gamma$-ray regime (e.g., \cite{vm95}).
Flares also have been observed at X- and gamma-ray bands by
multiwavelength observations of 
Mrk 421 (e.g., Macomb et al. 1995; Macomb et al. 1996
for erratum; Buckley et al. 1996)
and Mrk 501 (Catanese et al. 1997; Pian et al. 1998).
The tremendous luminosity and fast time variabilities from blazars
have led to the usual arguments that relativistic motion is occurring
in the emitting plasma. Moreover, the favored scenario to explain
these sources is that we are viewing nearly along the axis of a 
relativistically outflowing plasma jet that has been ejected
from an accreting super massive black hole (e.g., \cite{br78}). 

Although the origin of these multiwavelength spectra is still
under debate, several models on the radiative processes have
been put forth, in particular, models of Compton scattering of 
synchrotron photons or
external photons have been developed in recent years
(e.g., Bloom \& Marscher 1996;
Inoue \& Takahara 1996;
Ghisellini \& Madau 1996;
Dermer, Sturner, \& Schlickeiser 1997;
Mastichiadis \& Kirk 1997;
Sikora et al. 1997; 
B\"{o}ttcher, Mause, \& Schlickeiser 1997;
Georganopoulos \& Marscher 1998;
Ghisellini et al. 1998). 
Most of these calculations are either semi-analytic, or
for steady state situations, or not including the Compton
scattering process self-consistently. The main purpose of
this paper is to improve upon this situation. 
The physics of how energy is dissipated into relativistic
particles is, unfortunately, not well understood (see, however,
\cite{rl97}) and will not be treated fully in this paper.

Among various blazar models, synchrotron self-Compton (SSC) 
models have received a fair amount of attention,
by virtue of its simplicity and its possible predictive power. 
In these models, it is proposed that the nonthermal
synchrotron emission forms the radio-through-X-ray continuum, and
that the Compton scattering of these (soft) synchrotron photons by the
same nonthermal electrons produces the gamma rays ($\sim$ GeV -- TeVs).
In this paper, we focus on the so-called homogeneous SSC model
where a spherical blob of uniform relativistic plasma is postulated.
Even with such a greatly simplified picture, a number of parameters
have to be invoked, whose interplay gives rise to a rich dynamic 
behavior of the observed radiation. Of particular interest is
the correlated variabilities in X-ray and $\gamma$-ray fluxes,
since they represent the tail of nonthermal electrons which have
the shortest cooling timescale.
Although the generic multiwavelength spectra from radio to TeV 
can be fitted by a steady state model with fixed parameters 
(e.g, Kataoka et al. 1999), time-dependent calculations almost
always offer stronger constraints. 
Furthermore, when the self-Compton component contains a comparable
or even larger fraction of the radiative energy than the synchrotron
component, the whole problem becomes inherently nonlinear and
both components need to be calculated simultaneously and 
self-consistently. This naturally leads to the need of solving
coupled, time-dependent, nonlinear particle and photon 
kinetic-equations. 
Moreover, by examining the energy-dependence of flare data
at gamma-ray energies, one could potentially discriminate 
between SSC and external Compton-scattering
origins of the seed photons (Dermer 1998).

The simplest model for time variability of blazars 
(Mastichiadis \& Kirk 1997; hereafter MK97)
assumes that electrons obeying 
a power-law distribution are injected uniformly throughout
a relativistically moving blob over an extended period of time,
and that electrons cool by both synchrotron radiation 
and Compton scattering.
The blob is assumed not to accelerate or decelerate,
and the energy loss by Compton scattering of photons
impinging from outside the blob is assumed to be small 
in comparison with the synchrotron self-Compton loss.
MK97 reproduced the qualitative behavior of the energy-dependent
lags and the hysteresis diagrams (Takahashi et al. 1996).
Much of the work presented here follows closely to the previous
study by MK97, but we are using a completely different 
kinetic code which will be
discussed in later sections. 
Kirk, Rieger, \& Mastichiadis (1998) further modeled 
the evolution of synchrotron emission, 
calculating acceleration and cooling regions separately,
though Compton scattering was not included.

In this paper we present a detailed study of the time-evolution
of an electron-photon plasma (the positive particles could be either
protons or positrons) by solving the kinetic equations numerically. 
We briefly describe our model in \S \ref{sec:codes} and
show numerical results in \S \ref{sec:results}.
Summary is given in \S \ref{sec:sum}.

\section{MODEL}
\label{sec:codes}


We assume that observed photons are emitted from
a blob moving relativistically towards us
with a Doppler factor ${\cal D} = [\Gamma (1-\beta_{\Gamma} 
\cos\theta)]^{-1}$,
where $\Gamma$ is the Lorentz factor of the blob,
$\beta_{\Gamma} c$ is the velocity with $c$ being the light speed,
and $\theta$ is the angle between the direction of blob motion 
and the line of sight. 
The blob is a spherical and uniform cloud with radius $R$.
Relativistic electrons 
are injected into the blob and produce high energy emission.
Electrons and photons are uniformly distributed
throughout the blob.
The most important physical processes for electrons include 
synchrotron radiation and Compton scattering.
The spectra of electrons and photons in the blob are
calculated by solving the kinetic equations described below
(see also \cite{cb90}).

The kinetic equation describing the time-evolution of electron
distribution $n(\gamma)$
is given by
\begin{equation}
\label{eq:elkinetic}
\frac{\partial n}{\partial t} 
= - \frac{\partial}{\partial \gamma}
\left[ \left( \frac{d\gamma}{dt} \right)_{\rm loss}  n \right]
+ \frac{1}{2} \, \frac{\partial^2}{\partial \gamma^2} 
( D_e \, n ) 
 - \frac{n}{t_{e, {\rm esc}}} + Q(\gamma) \, ,
\end{equation}
where $n$ is the electron number density per $\gamma$,
$\gamma$ is the electron Lorentz factor, and $t_{e, {\rm esc}}$
is the time for electrons to escape from the blob.
The term $(d\gamma/dt)_{\rm loss}$ represents various
electron energy loss processes, such as synchrotron 
and Compton scattering; their corresponding energy diffusion
is given as $D_e$.  
We also include the synchrotron ``boiler'' effect (Ghisellini et al. 1988)
and other processes such as Coulomb collisions, 
though they are not important in the present 
study. Pair-production and annihilation are not treated in
the present code, though they tend to be less important too.
In our code, the particle equation is actually discretized in 
the momentum space so that thermal particles and their processes can be 
handled accurately. This is less important for AGN jet
parameters but will be very useful for modeling the emissions from 
stellar-mass black hole systems.
Note that equation (\ref{eq:elkinetic}) assumes a continuous 
electron energy loss (i.e., Fokker-Planck type). 
This assumption does not account for the situation when there is
a significant energy loss in a single
Compton scattering which is important in the Klein-Nishina regime.
This discrete nature of the Compton energy loss is, however, 
included in the photon kinetic equation, i.e., 
equations (\ref{eq:phkinetic}) and (\ref{eq:comp}).  
We have checked the accuracy of the continuous
approximation of the Compton energy loss in equation (1) in view of
energy conservation.  Our numerical tests show that the total energy
is conserved with the accuracy better than 5 per cent after 10 $R/c$,
when the SSC component is dominant and the scatterings occur frequently
in the Klein-Nishina regime.

The relevant kinetic equation for the time-evolution of
photons is given by
\begin{equation}
\label{eq:phkinetic}
\frac{\partial n_{\rm ph}(\epsilon)}{\partial t} 
= \dot{n}_{\rm C}(\epsilon) 
+ \dot{n}_{\rm em}(\epsilon) - \dot{n}_{\rm abs}(\epsilon) 
- \frac{n_{\rm ph}(\epsilon)}{t_{\rm ph, esc}} \, ,
\end{equation}
where $n_{\rm ph}(\epsilon)$ is the photon number
spectrum per unit volume per unit energy $\epsilon$.
Compton scattering is calculated as
\begin{equation}
\label{eq:comp}
\dot{n}_{\rm C}(\epsilon) 
=  - n_{\rm ph}(\epsilon) \, \int d\gamma \, n(\gamma) \, 
R_{\rm C}(\epsilon, \gamma) + \int\int d\epsilon^{\prime} \, d\gamma \, 
P(\epsilon; \epsilon^{\prime}, \gamma) \, 
R_{\rm C}(\epsilon^{\prime}, \gamma) \,
n_{\rm ph}(\epsilon^{\prime}) n(\gamma) \, .
\end{equation}
First term of the right hand side of equation (\ref{eq:comp}) denotes 
the rate that photons with energy $\epsilon$
are scattered by electrons with number spectrum $n(\gamma)$ 
per unit volume per unit $\gamma$;
$R_{\rm C}$ is the angle-averaged scattering rate.
Second term of equation (\ref{eq:comp}) denotes 
the spectrum of scattered photons:
$P(\epsilon; \epsilon^{\prime}, \gamma)$ is the probability 
that a photon with energy $\epsilon^\prime$ is scattered off 
by an electron with $\gamma$ to have energy $\epsilon$.
The probability $P$ is normalized such that
$\int P(\epsilon; \epsilon^{\prime}, \gamma) \, d\epsilon = 1$.
The details of $R_{\rm C}$ and $P$ are given in \cite{jones68}
and the appendix A of Coppi \& Blandford (1990).
We use the exact Klein-Nishina cross section 
in the calculations of Compton scattering.
Photon production and self-absorption by synchrotron radiation
are included in
$\dot{n}_{\rm em}(\epsilon)$ and $\dot{n}_{\rm abs}(\epsilon)$,
respectively.
The synchrotron emissivity and the absorption coefficient are calculated 
using the approximations given in Robinson and Melrose (1984) 
for transrelativistic electrons and Crusius and Schlickeiser (1986) 
for relativistic electrons.
External photon sources such as disk photons can be included, 
though they are not considered here.
The rate of photon escape is estimated as 
$n_{\rm ph}(\epsilon)/t_{\rm ph, esc}$.
Since we are in the optically thin limit, 
we set $t_{\rm ph, esc} = R/c$,
which is a good approximation.
The photon spectra from 
solving equation (\ref{eq:phkinetic}) has been extensively 
compared with those from Monte-Carlo simulations and we have
found very good agreement between them (Kusunose, Coppi, \& Li 1999).

The comoving quantities are transformed back 
into the observer's frame using
$\epsilon_{\rm obs} = \epsilon \, {\cal D}/(1+z)$
and $dt_{\rm obs} = dt \, (1+z)/{\cal D}$, where
$z$ is the redshift of the source.

We assume that electrons are injected obeying a power law in energy:
$$Q(\gamma, t) = Q_0(t) \, \gamma^{-p} $$ 
with $\gamma_{\rm min} \leq \gamma \leq \gamma_{\rm max}$.
The total energetics of the electrons can be represented
by a compactness parameter which is proportional to $L/R$,
where $L$ is the source luminosity. We do not consider specific
acceleration mechanisms in this paper.  Thus particles are just
being injected into the blob over a finite time. We emphasize
that in order to be consistent with the basic assumption of
spatial homogeneity, we require the injection time $t_{\rm inj}$
to be longer than $t_{\rm dyn} = R/c$. In effect, we can not
probe variabilities shorter than $t_{\rm dyn}$ in the comoving frame.
Other physical effects such as adiabatic loss via expansion
might play an important role but is not considered here.
It is unfortunate that we need such a large number of parameters
to proceed with the calculations, and this is the main reason
we opt not to add further complications such as acceleration.

\section{RESULTS}
\label{sec:results}

The dynamic behavior of the emission spectra is controlled by
several timescales, namely, the cooling time $t_{\rm cool}$,
the dynamic time $t_{\rm dyn}$, the injection duration $t_{\rm inj}$,
and the escape time $t_{\rm esc}$. The causality argument requires
that $t_{\rm dyn} \leq t_{\rm inj}, t_{\rm esc}$, whereas $t_{\rm cool}$
can be smaller than $t_{\rm dyn}$.  For SSC models, both
synchrotron and Compton processes contribute to the electron cooling,
so $1/t_{\rm cool} = 1/t_{\rm syn} + 1/t_{\rm ssc}$, where
$t_{\rm syn} = \gamma/|{\dot \gamma}_{\rm syn}|$ and
$t_{\rm ssc} = \gamma/|{\dot \gamma}_{\rm ssc}|$.

In the following analyses, we will divide our results into two 
major parts, based on whether $t_{\rm cool}$ is longer or shorter
than $t_{\rm inj}$. In the limit of $t_{\rm cool}
\geq t_{\rm inj} \geq t_{\rm dyn}$, where $t_{\rm cool}$ is evaluated
using the highest particle energy, the injected particle distribution
does not change appreciably during the injection process. We call this
the injection-dominated limit. On the
other hand, if $t_{\rm inj} \geq t_{\rm dyn} \geq  t_{\rm cool}$,
then particles will be sufficiently cooled while the injection still
occurs, and emissions are from a cooled particle distribution
rather than from the injected one. We call this the 
cooling-dominated limit. Consequently, we expect rapid variations
in both fluxes and spectra in the latter case and relatively slow
spectral variations in the former case.

The primary purpose of this paper is to understand the dynamics of
SSC model.  We thus have chosen a broad range of parameters rather than
try to fit any specific source spectrum, but we certainly use 
parameters thought to be applicable to those
sources (Mrk 421 in particular) to guide our calculations.

\subsection{Long Cooling Time Limit}

We show that, in this limit, a set of SSC model 
parameters can be uniquely determined from the observable quantities.
We use observations of Mrk 421 as an example and 
further discuss their implications for
multiwavelength observations.

\subsubsection{A Unique Solution}
\label{sec:us}

Let $\cal E$ represent the energy (in ergs) injected in nonthermal electrons,
which we assume can be described as $N_e(\gamma) = N_0 \, \gamma^{-p}$
and $\gamma_{\rm min} \leq \gamma \leq \gamma_{\rm max}$.
Then $N_0 = (2-p) {\cal E} / [m_e c^2 (\gamma_{\rm max}^{2-p}
- \gamma_{\rm min}^{2-p})]$ (for $p\ne 2$).
In the limit of long cooling time, we can use the injected
electron distribution to calculate the synchrotron and SSC fluxes.
Thus, for the peak energies of synchrotron and SSC, we have
\begin{equation}
\label{eq:nusyn}
{\cal D} \gamma_{\rm max}^2 B \approx \nu_{\rm syn}/2.8\times 10^6
= {\tilde \nu}_{\rm syn}~~,
\end{equation}
\begin{equation}
\label{eq:nussc}
{\cal D} \gamma_{\rm max} \approx \nu_{\rm ssc}/1.236\times 10^{20}
= {\tilde \nu}_{\rm ssc}~~,
\end{equation}
where $\nu_{\rm syn}$ and $\nu_{\rm ssc}$  are the
synchrotron and $\gamma$-ray peaks in Hz, respectively, and
$B$ is magnetic fields in Gauss. The numerical normalization factors
are easily obtainable using the standard expressions for the peak
synchrotron energy and inverse Compton peak energy in the KN limit.
For Mrk 421, we have ${\tilde \nu}_{\rm syn} \approx 1.72\times 10^{11}$
($\sim 2$ keV) and ${\tilde \nu}_{\rm ssc} \approx 10^7$
($\sim 5$ TeV).  From equations (\ref{eq:nusyn}) and (\ref{eq:nussc}),
we get
\begin{equation}
\label{eq:mag}
B = {\cal D} {\tilde \nu}_{\rm syn}/{\tilde \nu}^2_{\rm ssc}~~.
\end{equation}

The relative ratio $\eta$ of SSC to synchrotron fluxes for Mrk 421
is close to 1.  This ratio can be approximately represented by
the ratio of the comoving-frame synchrotron photon energy-density,
$U_{\rm syn} = L_{\rm syn}/(4\pi R^2 c {\cal D}^4)$, 
to the magnetic field energy-density, $U_B = B^2/(8 \pi)$. 
But the exact value of $\eta$ could be quite different from this
estimate owing to several factors: the end point effect
at the peak of both synchrotron and SSC fluxes,
\footnote{For $p < 3$, the peak of $\nu L_{\nu}$
is at $\nu_{\rm syn} \propto \gamma^2_{\rm max}B$, but its flux 
is {\em smaller} than that obtained using the $\delta-$function
approximation to the scattering cross section. To be consistent with
our numerical code results which have used the {\em exact} 
synchrotron emissivity formulations, we have
introduced a reduction factor of $f_{\rm syn} (< 1)$ in our simplified
analytic estimates for the peak synchrotron flux (see also
equation (\ref{eq:lsyn}). 
Note that this reduction only
applies to the end points of emissivity. For energies much smaller
than $\nu_{\rm syn}$, the $\delta-$function approximation is quite
accurate.}
the KN effect of Compton scattering for producing TeV emission, 
and the fact that
the electron distribution will be slightly cooled even though
the cooling timescale is long. It is very difficult to get an exact
analytic value to account for all these effects, so we introduce
a correction factor $f_c$ in calculating $\eta$.
Thus, we have
\begin{equation}
\label{eq:eta}
\eta = f_c~\frac{U_{\rm syn}}{U_B} ~~{\rm or}~~
L_{\rm syn} \approx {\cal D}^4 4\pi R^2 c ~ U_B~(\eta / f_c) \, .
\end{equation}
Furthermore, since all blazar sources are observed to be highly variable,
an additional constraint has usually been proposed that
\begin{equation}
R \approx {\cal D} ~c t_{\rm var}~~,
\end{equation}
where $t_{\rm var}$ is defined as a variability timescale
in the observer's frame. 

Combining all the equations given above, we have 4 equations for
4 independent variables (i.e., ${\cal D}, B, \gamma_{\rm max},$ and
$R$). The supplemental information include
${\tilde \nu}_{\rm syn}, {\tilde \nu}_{\rm ssc}, L_{\rm syn}$, $\eta$,
and $t_{\rm var}$, with a somewhat variable factor $f_c$. 
Thus we can {\em uniquely} determine a solution set for all the
parameters. The Doppler factor from this solution (denoted
by a subscript `$s$') can be expressed as
\begin{equation}
\label{eq:solution}
{\cal D}_s \approx 22.2 
\left(\frac{L_{\rm syn}}{6\times 10^{44}}\right)^{\frac{1}{8}}
\left(\frac{{\tilde \nu}_{\rm ssc}}{10^7}\right)^{\frac{1}{2}}
\left(\frac{{\tilde \nu}_{\rm syn}}{1.7\times 10^{11}}\right)^{-\frac{1}{4}}
\left(\frac{t_{\rm var}}{10^4}\right)^{-\frac{1}{4}}
\left(\frac{f_c}{0.4}\right)^{\frac{1}{8}}
\left(\frac{\eta}{1.0}\right)^{-\frac{1}{8}}~~.
\end{equation}

The peak luminosity ($L_{\rm syn}$) of the synchrotron component 
in the observer's frame can be estimated using $\nu L_{\nu}$ at
$\nu_{\rm syn}$ (again taking into account the end-point effects), 
\begin{equation}
\label{eq:lsyn}
L_{\rm syn} \approx {\cal D}^4~ f_{\rm syn}
\frac{2(2-p)}{3} \sigma_{\rm T} c U_B
\frac{{\cal E}}{m_e c^2}~\gamma_{\rm max}~~,
\end{equation}
where $\sigma_{\rm T}$ is the Thomson cross section,
$f_{\rm syn} (< 1)$ represents the reduction of synchrotron
flux at the end point, ${\cal D}^4$ is due to the Doppler boosting.
Observationally,
$L_{\rm syn}$ at $\sim 2$ keV is $\geq 6\times 10^{44}$ ergs s$^{-1}$
with a luminosity distance of $3.8 \times 10^{26}$ cm ($z = 0.0308$)
for a $q_0 = 1/2$ and $\Lambda = 0$ cosmology. Here, 
$H_0 = 75$ km s$^{-1}$ Mpc$^{-1}$ is assumed. Also, 
fitting of Mrk 421's synchrotron spectrum suggests $p \approx 1.65$
(MK97).  From this, we can determine the rest of the parameters for
this unique solution (again denoted by a subscript `s')
\begin{eqnarray}
B_s & = & {\cal D}_s {\tilde \nu}_{\rm syn}/ 
{\tilde \nu}_{\rm ssc}^2 
\approx 0.038~~ {\rm G}~~,\label{eq:b-d}\\
\gamma_{{\rm max}, s} & = &{\tilde \nu}_{\rm ssc} / {\cal D}_s
\approx 4.5\times 10^5~~,\\
R_s & =& c \, t_{\rm var} {\cal D}_s \approx 6.5\times 10^{15} 
~~{\rm cm}~~,\\
{\cal E}_s &=& \frac{1.55\times 10^9}{(2-p) f_{\rm syn}}~
\frac{L_{\rm syn}}{\gamma_{{\rm max}, s} B^2_s {\cal D}^4_s}
\approx 4.1 \times 10^{46} ~~{\rm ergs} ~~,\label{eq:e-d}
\end{eqnarray}
where we have used $p = 1.65$ and $f_{\rm syn} = 0.4$. 
\footnote{The fact that $f_c$ and $f_{\rm syn}$ are both chosen as
$0.4$ is a coincidence. The evaluation of $f_c$ involves end-point
effects from both synchrotron and Compton scattering, whereas 
$f_{\rm syn}$ is only concerned with synchrotron. The value
$0.4$ is obtained by comparing the analytic estimates with
the exact numerical calculations.}
More importantly, we can check our original assumption that
cooling time is long compared to $t_{\rm dyn} = R/c$.
This is obviously satisfied since 
$t_{\rm syn} \approx (6\pi m_e c/\sigma_{\rm T})/(\gamma_{\rm max} B^2)
\approx 1.2\times 10^6$ sec, which is much longer than 
$t_{\rm dyn}\approx 2.2 \times 10^5$ sec. 

\begin{figure}
\epsfig{file=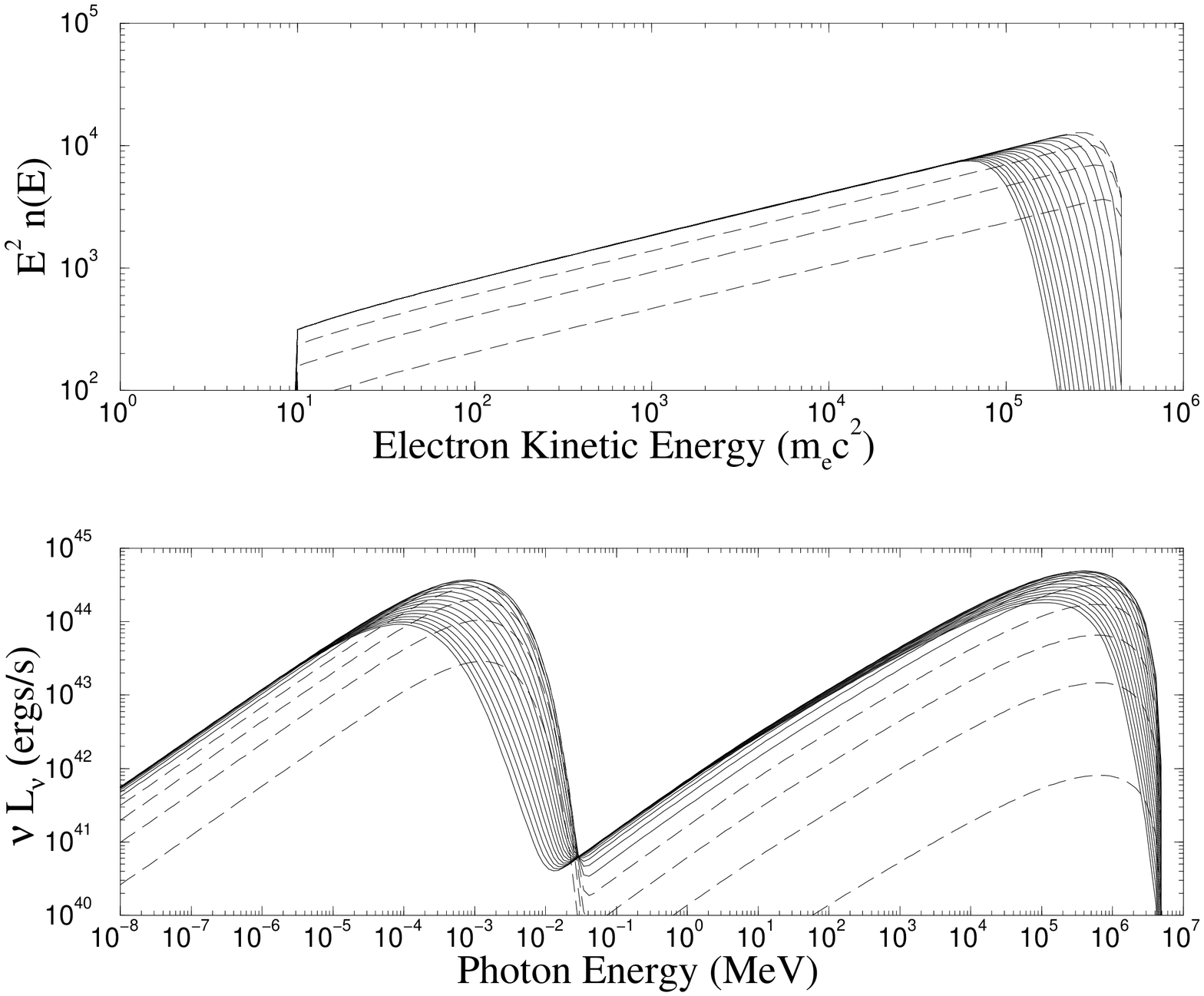,height=5in,width=4in,angle=-90}
\caption{Time evolution of electron energy distribution (upper panel)
and the corresponding photon spectra (lower panel) 
using parameters from the unique solution discussed in \S \ref{sec:us}.
Electrons are injected at a constant
rate lasting $2 t_{\rm dyn}$ in the comoving frame.
There are 20 curves in each panel, which start at $t = 0.5 t_{\rm dyn}$
and end at $t = 10 t_{\rm dyn}$  with a time interval of 
$0.5 t_{\rm dyn}$ between them. The injection process is shown by
the dashed curves moving up in $n(E)$, 
along with the increasing photon fluxes.
After the injection stops, electrons are continuously cooled and
the photon spectrum softens. These parameters give a comparable
peak fluxes for synchrotron and SSC components.
}
\label{fig:parpht-us}
\end{figure}

Using the above derived parameters, we solve 
equations (\ref{eq:elkinetic}) and (\ref{eq:phkinetic})
simultaneously and follow the evolution until 
$20 t_{\rm dyn}$. A total energy of ${\cal E}$ is injected
in nonthermal electrons over a comoving timescale of
$t_{\rm inj} = 2 t_{\rm dyn}$.
In these calculations electrons are not allowed to escape.
Figure \ref{fig:parpht-us} shows the time evolution of electron and
photon distributions as the system evolves.
Note that the time for synchrotron and SSC components to
reach their peak fluxes is different, and that it happens
after the electron injection has stopped.
To qualitatively understand this, we can write the photon 
kinetic equation symbolically as
\begin{equation}
\label{eq:dndt}
\frac{\partial n_{\rm ph}(\epsilon)}{\partial t} 
 = {\rm Production}(\epsilon) - {\rm Escape}(\epsilon)~~.
\end{equation}
Thus, the photon flux at energy $\epsilon$ will increase
if the production rate is larger than the escape and decrease
if escape is quicker. This determines when the peak of photon
flux at certain energy is reached. Using photon flux at keV
as an example, the production of these photons still continues
even when the electron injection stops because the cooling timescale
is much longer than $R/c$. Eventually as electrons cool, they
can no longer produce keV synchrotron emissions, the flux
at keV starts to decline.

\begin{figure}
\epsfig{file=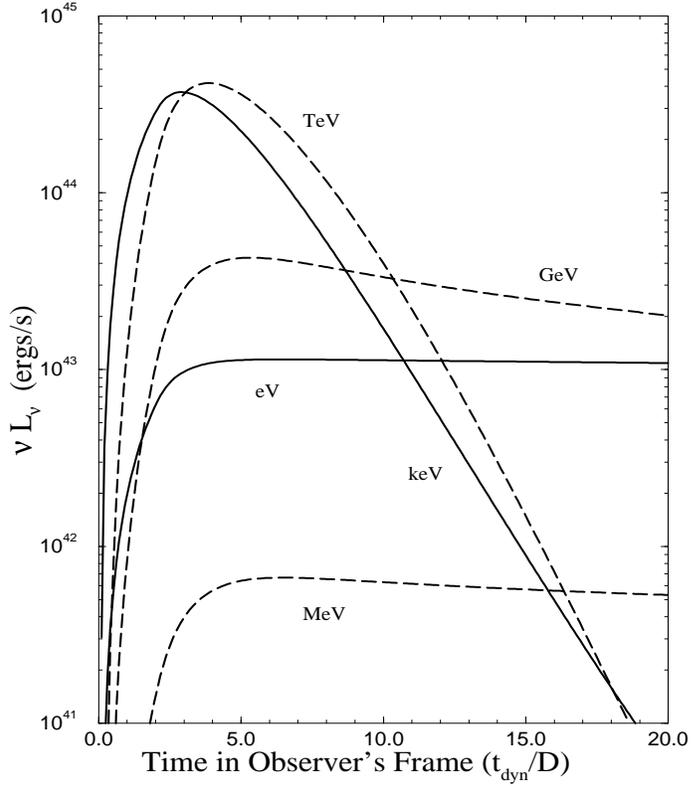,height=5in,width=4in,angle=0}
\caption{Multiwavelength light curves in observer's frame
(${\cal D}$ is assumed to be 10) using parameters from
the unique solution (cf. Figure \ref{fig:parpht-us}).
Fluxes at the synchrotron and SSC peaks show fast time variability
with large amplitudes, but fluxes at other wavelengths have
a very long plateau with very small amplitude variation.
}
\label{fig:ltcv-us}
\end{figure}

Figure \ref{fig:ltcv-us} shows the light curves at different energy bands
expected from this injection event. Since the cooling time
is rather long, only the energy bands corresponding to the tail
of electron distributions show short time variabilities
due to that electron injection is turned on and off;
whereas other energy bands show a long plateau, representing
a balance between the photon production and escape.
A clear prediction from this is that there should be very 
little spectral evolution except at the peaks of synchrotron 
and SSC. All these are commensurate with the dynamics 
of electron cooling.

\begin{figure}
\epsfig{file=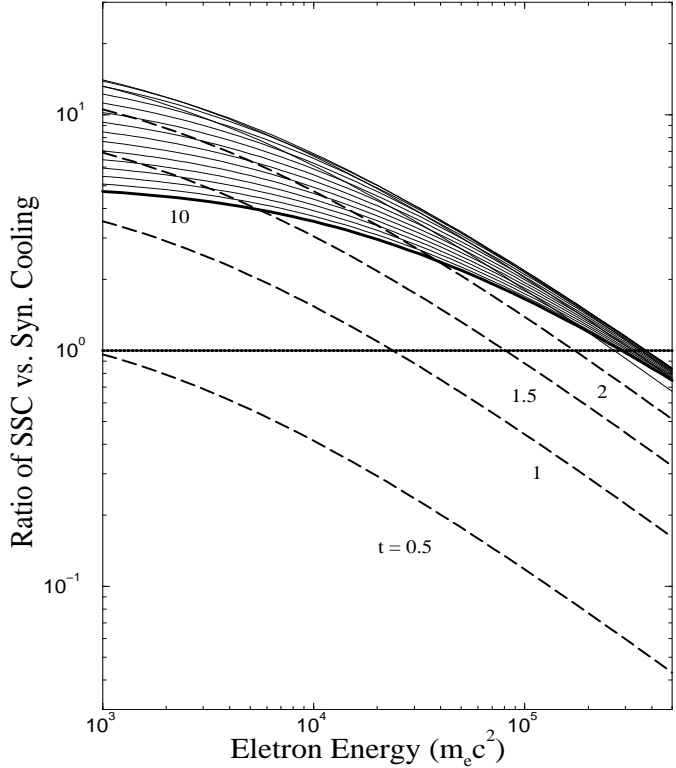,height=5in,width=4in,angle=0}
\caption{Shown is $|\dot\gamma_{\rm ssc}/\dot\gamma_{\rm syn}|$
as a function of electron energy at different times 
(from $0.5 - 10 t_{\rm dyn}$)
using parameters from the unique solution 
(cf. Figure \ref{fig:parpht-us}). The horizontal solid line at 
the ratio being $1$ is plotted to guide the comparison.
The SSC process becomes important as soon as the photon energy
is built up and, in fact, is more important than synchrotron cooling
for most of the electron energies.
}
\label{fig:syn-ssc-ratio}
\end{figure}

In order to further differentiate the role of synchrotron
versus Compton cooling on electrons, we plot the ratio
of $|\dot\gamma_{\rm ssc}/\dot\gamma_{\rm syn}|$ as a
function of electron energies at different times in Figure 
\ref{fig:syn-ssc-ratio}.
It is clear that SSC cooling becomes more important than
synchrotron cooling when the photon energy density builds up 
within the system as time proceeds. After reaching the peak,
the SSC cooling starts to decrease as the photon energy density
decreases. The dependence of this ratio on the electron energy
is partly due to the KN effect. This figure clearly indicates that
one can not ignore SSC cooling in estimating certain parameters
and that the evolution is very nonlinear, thus a self-consistent
numerical calculation is required for fitting
the data more accurately.

Note that the spectra given in Figure \ref{fig:parpht-us} is not intended to
be an accurate fit to the observed spectra from Mrk 421.
In fact, the TeV spectrum of Mrk 421 is known to be roughly
a power law (\cite{ketal99}), but the generic spectrum in TeV obtained here
has a clear curvature, due to the fact that the Compton scattering
is in the KN regime. Nevertheless, this exercise allows us to
establish a parameter space where a reasonable fit to the
actual spectrum might be obtainable, and it has the nice feature
that the electron energy distribution retains the injected form
without much softening, which greatly simplifies the analysis.

\subsubsection{Parameter Variations on the Unique Solution}

\begin{figure}
\epsfig{file=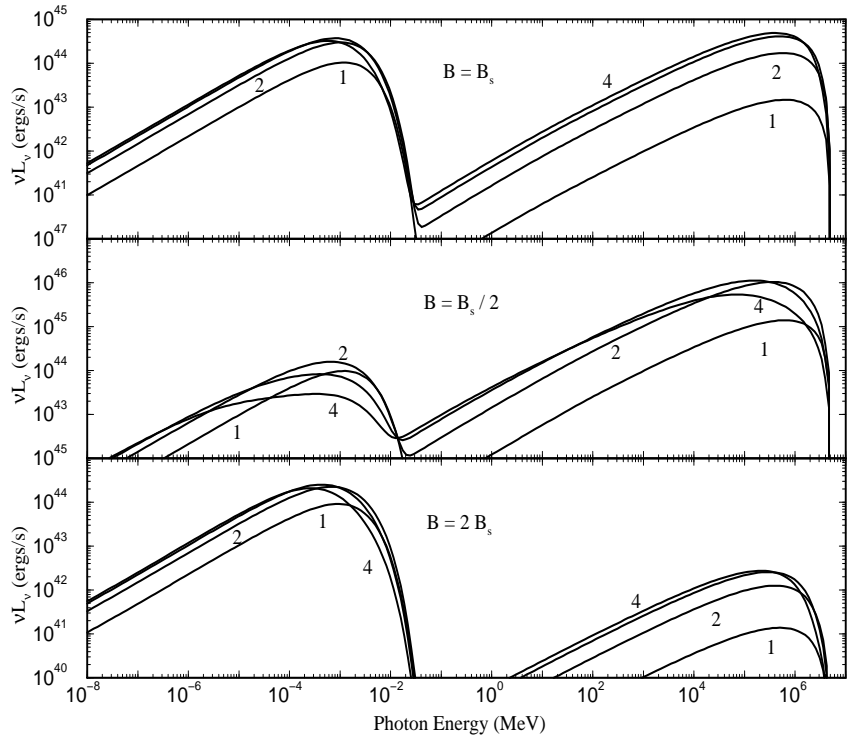,height=5in,width=4in,angle=-90}
\caption{Shown is the photon spectral evolution from $0 - 4 t_{\rm dyn}$
as the magnetic field is varied from the unique solution value $B_s$.
The ratio $\eta$ is unity when $B = B_s$ (cf. Figure \ref{fig:parpht-us})
but varies nearly as $(B_s/B)^8$ as shown in the middle and lower
panels here.
}
\label{fig:varyb}
\end{figure}

In this subsection, we explore how sensitive the above results
are to parameter variations. The parameter we want to emphasize
is the ratio $\eta$ of the SSC component to 
the synchrotron component.  From equations (\ref{eq:mag})
and (\ref{eq:solution}),
and holding other parameters unchanged, we can see that 
\begin{equation}
\label{eq:eta-b}
\eta \propto {\cal D}^{-8} \propto B^{-8}~~,
\end{equation}
which implies that a small change in magnetic field and/or Doppler
factor could result in a large variation in $\eta$. 
Figure \ref{fig:varyb} shows this effect when $B$ is being varied.
The general trend from equation (\ref{eq:eta-b}) is indeed
confirmed, i.e., smaller/larger $B$ (versus $B_s$) gives 
much larger/smaller $\eta$. For $B < B_s$, the variation amplitude
in $\eta$ does not exactly follow equation (\ref{eq:eta-b}).
We attribute this discrepancy mostly to variation in 
$f_c$ because both $\gamma_{\rm max}$ and synchrotron
photon energy in comoving frame are varying for different $B$.
Additionally, when $\eta > 1$ (SSC cooling is more important
than synchrotron cooling), electron distribution is subject
to stronger cooling. This is evident in its rapid spectral
variation in the middle panel of Figure \ref{fig:varyb}. 
For larger $B$ which results in $\eta < 1$,
equation (\ref{eq:eta-b}) is mostly confirmed. Note that
the Doppler factor ${\cal D}$,
using equation (\ref{eq:mag}), is now getting uncomfortably large.

To conclude, in the limit that the cooling time of the highest 
energy electrons
is longer than the dynamic timescale on which injection occurs,
we can most likely find a {\em unique} set of parameters that 
roughly satisfy
the observational constraints. The most important prediction is
that even though the fluxes of the synchrotron and SSC peaks
can vary by a large factor ($> 10$) in a short time whose time
histories look like ``pulses'', the duration of emission at other wave bands 
(such as GeV, MeV, eV) can be considerably longer by at least
a factor of $4$, which is commensurate with long electron cooling
time at those energies. 

This parameter space has some interesting implications
for interpreting multiwavelength data from blazar monitoring campaigns.
Using keV and eV bands as an example (similar arguments can be
made for TeV and GeV bands too),
since the lifetimes for keV-producing and eV-producing electrons are 
different, there is really no reason to expect their light curves
to track each other and to have the same rise and fall patterns
or timescales. Furthermore, when there are multiple injections
occurring over a timescale shorter than the lifetime of
keV-producing electrons but longer than the lifetime of 
eV-producing electrons, fluxes at keV could vary rapidly to
reflect the multiple injections. 
Fluxes at eV, however, might only show a 
continuous increase with no obvious decline because of 
the accumulation of eV-producing electrons from multiple injections 
(see also \S \ref{sec:multi}).

\subsection{Short Cooling Time Limit}

In this subsection we explore the limit where
$t_{\rm cool}(\gamma_{\rm max})
< t_{\rm dyn}$, i.e., electrons are efficiently being cooled 
while they are injected into the system. 
All the runs in this section have
size $R = 1.5 \times 10^{16}$ cm which gives a comoving
dynamic timescale $R/c = 5\times 10^5$ sec,
particle's $\gamma_{\rm min} = 10$, $\gamma_{\rm max} = 10^{6}$,
and index $p=2$. The Doppler factor is chosen as 10 and magnetic
field $B = 0.1$ G, though most of the conclusions depend
weakly on ${\cal D}$ and $B$ in this section. 
The particularly
attractive feature of this limit is that it is possible to
achieve short time variabilities for many wave bands, contrary
to the case shown in Figure \ref{fig:ltcv-us} from the previous section.

One serious problem in comparing the theoretical results with
the actual observations is that most observations need to 
accumulate over certain time interval
(to collect enough photons) and different
integration times are needed for different energy bands. 
So, in order to make a direct comparison with the observations,
properly averaged fluxes are  needed,
rather than the prompt flux we have presented above.
This inevitably introduces many more additional parameters
in determining how photons at different energy bands are sampled.
To avoid these complications, as before, 
we will continue to present the prompt
flux results, leaving the problem of integrated fluxes
to future studies on detailed spectral fitting of 
particular sources.

To quantify the relative importance of SSC versus synchrotron,
the previous expression for $\eta$ has to be modified.
To recapitulate, the comoving synchrotron photon energy-density is
$U_{\rm syn} \simeq [ m_e c^2 / ( 4 \pi R^2 c)] 
\int_{\gamma_{\rm min}}^{\gamma_{\rm max}} N_e(\gamma) | \dot{\gamma} |
\, d \gamma$,
where the electron energy loss-rate through synchrotron radiation
is given by $\dot{\gamma} = - [ 4 \sigma_{\rm T} / (3 m_e c) ]
U_B \gamma^2$. Thus,  we get
\begin{equation}
\label{eq:ussc1}
\eta = f_c~\frac{U_{\rm syn}}{U_B} 
= f_c~\frac{\sigma_{\rm T}}{3\pi R^2} \, \frac{{\cal E}}{m_e c^2}
\frac{2-p}{3-p} \,
\frac{\gamma_{\rm max}^{3-p} - \gamma_{\rm min}^{3-p}}
{\gamma_{\rm max}^{2-p} - \gamma_{\rm min}^{2-p}}
\approx f_c~ \frac{\sigma_{\rm T}}{3\pi R^2} \,
\frac{{\cal E}}{m_e c^2} \frac{\gamma_{\rm max}}
{\ln (\gamma_{\rm max} / \gamma_{\rm min})} \, ,
\end{equation}
where the last expression applies to the case $p = 2$.
On the other hand, if $t_{\rm dyn} > t_{\rm cool}$, a cooled
electron distribution has to be used when calculating the
synchrotron photon energy-density. A very rough estimate of
the electron break energy $\gamma_{\rm br}$,
beyond which cooling dominates can be given as 
$$\frac{6\pi m_e c}{\sigma_{\rm T}}~\frac{1}{\gamma_{\rm br} B^2} \approx 
{\rm max}(t_{\rm inj}, t_{\rm dyn})~~.$$
The cooled electron distribution has the original index $-p$ between
$\gamma_{\rm min}$ and $\gamma_{\rm br}$, and roughly $-p-1$ between 
$\gamma_{\rm br}$ and $\gamma_{\rm max}$. Since the number density
of electron at high energy end is very small for $p > 1$, the
previous expression for $N_0$ (\S\ref{sec:us}) still applies. 
Then we can derive another expression for $\eta$ as
\begin{equation}
\label{eq:ussc2}
\eta_c = f_c~\frac{\sigma_{\rm T}}{3\pi R^2} \frac{\cal E}{m_e c^2}\,
 \left[\left(\frac{2-p}{3-p}\right)
\frac{\gamma^{3-p}_{\rm br}}{\gamma^{2-p}_{\rm max}}
~ +~ 1 \right]\,
\approx f_c~\frac{\sigma_{\rm T}}{3\pi R^2} \frac{\cal E}{m_e c^2}\,
\left[\frac{\gamma_{\rm br}~+~\ln (\gamma_{\rm max} / \gamma_{\rm br})}
{\ln (\gamma_{\rm max} / \gamma_{\rm min})}\right] \, ,
\end{equation}
where again, the last expression applies to the case $p = 2$.
Note that this ratio $\eta_c$ depends on $B$ through $\gamma_{\rm br}$.
Using the same parameters given previously, we find that
equations (\ref{eq:ussc1}) and (\ref{eq:ussc2}) give
$\eta_c \simeq (\gamma_{\rm br}/\gamma_{\rm max}) \eta$.
In other words, in order to reach the same relative ratio
between synchrotron and SSC,
more energy is needed (by a factor of 
$\gamma_{\rm max}/\gamma_{\rm br}$ for $p=2$) 
if electrons are cooled efficiently during injection.

\subsubsection{Dynamics of a Single Injection}

In this subsection we concentrate on the dynamics of a single injection
event lasting $t_{\rm inj}$ and its corresponding
evolution of electron and photon distributions.
The idea is to mimic individual flaring events in blazars
and gain some basic knowledge of how synchrotron and SSC
components are dynamically linked.
To simplify the calculations and analysis,
we choose 6 total injection energies
with a factor of 10 increase from $10^{44}$ ergs to $10^{49}$ ergs.
These energies in nonthermal electrons are injected 
over a comoving timescale of $t_{\rm inj} = 2 t_{\rm dyn}$.
We solve equations (\ref{eq:elkinetic}) and (\ref{eq:phkinetic})
simultaneously and follow the evolution until 
$10 t_{\rm dyn}$. Electrons are not allowed to escape.
With these parameters, $t_{\rm cool}(\gamma_{\rm max})$ is
much smaller than $t_{\rm dyn}$, so electrons are appreciably
cooled during injection. 

\begin{figure}
\epsfig{file=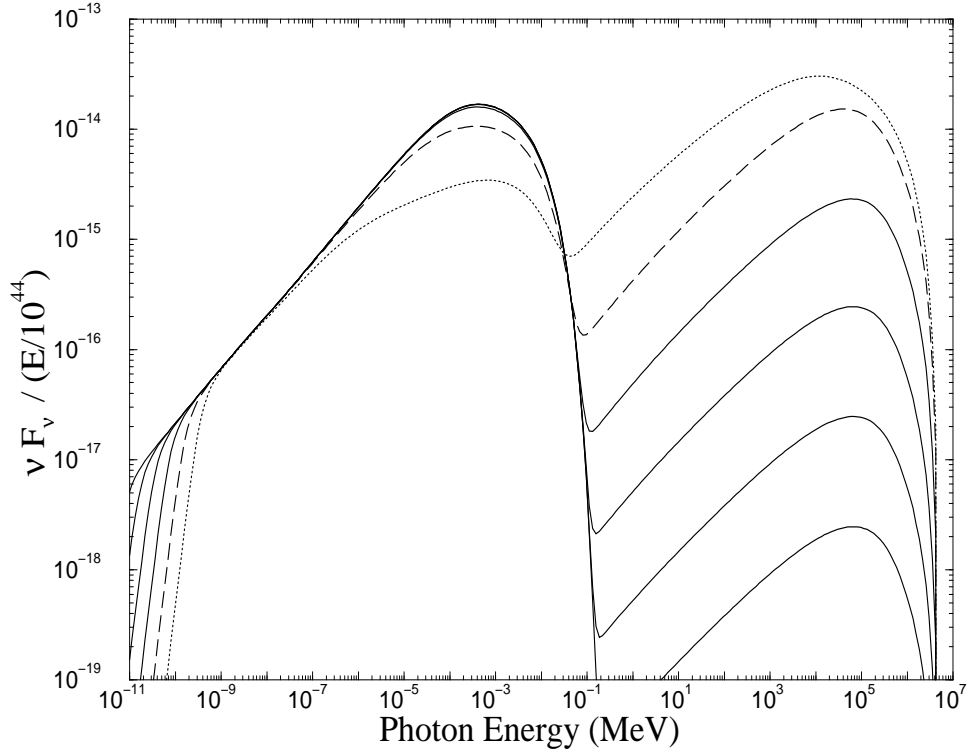,height=5in,width=4in,angle=-90}
\caption{The photon spectra $\nu F_\nu$ for 6 different 
total-injection-energy ${\cal E}$, ranging from $10^{44}$ ergs to
$10^{49}$ ergs with a factor of 10 increment in each case.
All 6 spectra are taken at the end of electron injection 
($t = 2 t_{\rm dyn}$) and their fluxes are divided by 
${\cal E}/10^{44}$ so that if the synchrotron flux was 
exactly proportional to ${\cal E}$, they would have had the 
same heights. As ${\cal E}$ increases, the electron cooling
changes from synchrotron dominated (when ${\cal E} \leq 10^{47}$)
to synchrotron self-Compton (SSC) dominated. Note that the
increase in SSC component is proportional to ${\cal E}^2$ 
for ${\cal E} \leq 10^{47}$. When SSC cooling is very strong,
electrons cool so quickly that the synchrotron flux at
$2 t_{\rm dyn}$ is no longer scaled as ${\cal E}$ anymore
as shown in the $10^{48}$ ({\it dashed}) 
and $10^{49}$ ({\it dotted}) cases.  Also in these
large ${\cal E}$ cases, the efficient cooling makes the 
synchrotron peak rather broad.
}
\label{fig:6curves}
\end{figure}

Figure \ref{fig:6curves} shows the $\nu F_\nu$ spectra of these
6 injections taken at $t = t_{\rm inj}$ and with fluxes 
divided by ${\cal E}/10^{44}$.
It is evident that synchrotron is the dominant cooling process
for ${\cal E} \leq 10^{47}$ cases, whereas SSC becomes the dominant 
cooling process as the photon energy density builds up in the
$10^{48}$ and $10^{49}$ cases. In fact, electron cooling is so
significant in the $10^{49}$ case that the maximum synchrotron flux
is reduced by almost a factor of 10 compared to other lower 
injection energy cases, and its SSC peak energy is also much 
softer than others.

As given in equation (\ref{eq:ussc2}), the ratio of SSC to
synchrotron is roughly proportional to ${\cal E}$. Thus as ${\cal E}$ 
increases, so is this ratio. This is shown as the (almost) linear
increase in the SSC peaks of Figure \ref{fig:6curves}. So we
conclude that so long as the peak in SSC is less than synchrotron,
the magnitude of increase in SSC peak will be the {\em square} of
the magnitude of increase in synchrotron peak. When the 
SSC component becomes comparable to synchrotron component, the 
system becomes highly nonlinear, the estimate of $\gamma_{\rm br}$
based on pure synchrotron cooling is no longer valid, even though
equation (\ref{eq:ussc2}) probably still applies as long as a
cooled electron distribution is used. 

A further point regarding the relative ratio of SSC versus synchrotron
components is that the initial $\sim 10$ GeV -- TeV production via SSC
is in the KN regime, which reduces the SSC flux. 
This implies that an even larger ${\cal E}$ is needed than those 
given in equation (\ref{eq:ussc2}). 
Observations of Mrk 421 and 501 seem to indicate roughly the same
heights of synchrotron and SSC peaks. Thus, in modeling the 
time-dependent (or even steady state) emissions from these objects,
a full KN cross section has to be used, as was done here in our code.

\subsubsection{Dynamics and Light Curves}

\begin{figure}
\epsfig{file=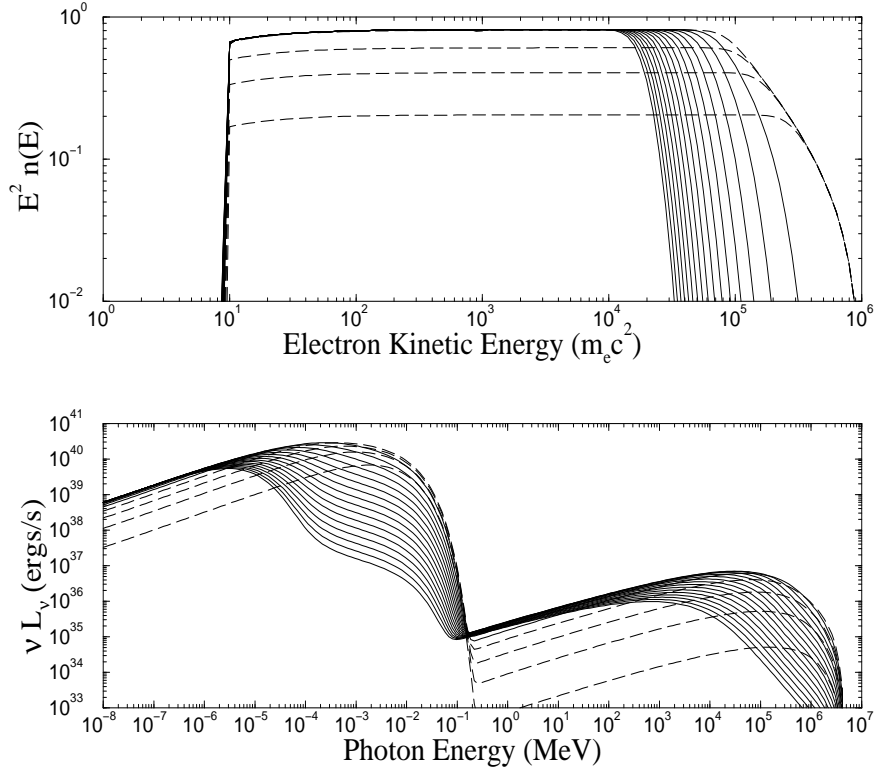,height=5in,width=4in,angle=-90}
\caption{Time evolution of electron energy distribution (upper panel)
and the corresponding photon spectra (lower panel) for
${\cal E} = 10^{44}$ ergs. Electrons are injected at a constant
rate lasting $2 t_{\rm dyn}$ in the comoving frame, with 
$\gamma_{\rm min} = 10$, $\gamma_{\rm max} = 10^6$, and index $p = 2$.
There are 20 curves in each panel, which start at $t = 0.5 t_{\rm dyn}$
and end at $t = 10 t_{\rm dyn}$  with a time interval of 
$0.5 t_{\rm dyn}$ between them. The injection process is shown by
the dashed curves moving up in $n(E)$, 
along with the increasing photon fluxes.
After the injection stops, electrons are continuously cooled and
photon spectrum softens (solid curves).  With this injection energy,
the SSC peak is considerably lower than the synchrotron peak.}
\label{fig:pp-e44}
\end{figure}

\begin{figure}
\epsfig{file=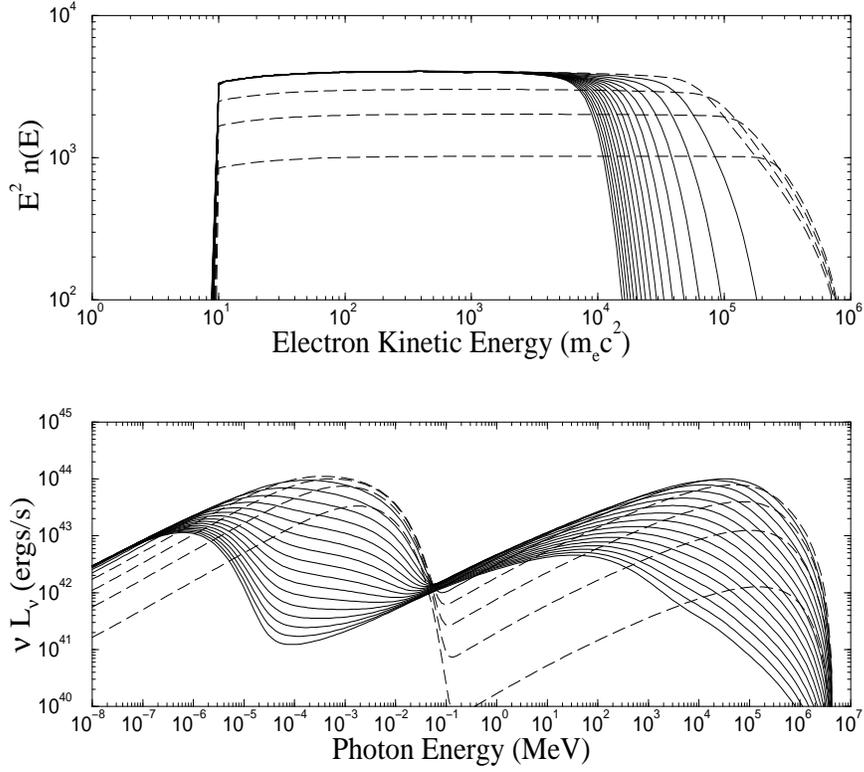,height=5in,width=4in,angle=-90}
\caption{Same as Figure \ref{fig:pp-e44}, except that
${\cal E} = 5\times 10^{47}$ ergs. Now the SSC component has become
comparable to the synchrotron component, and the whole system
understandably evolves on a faster timescale.}
\label{fig:pp-5e47}
\end{figure}

\begin{figure}
\epsfig{file=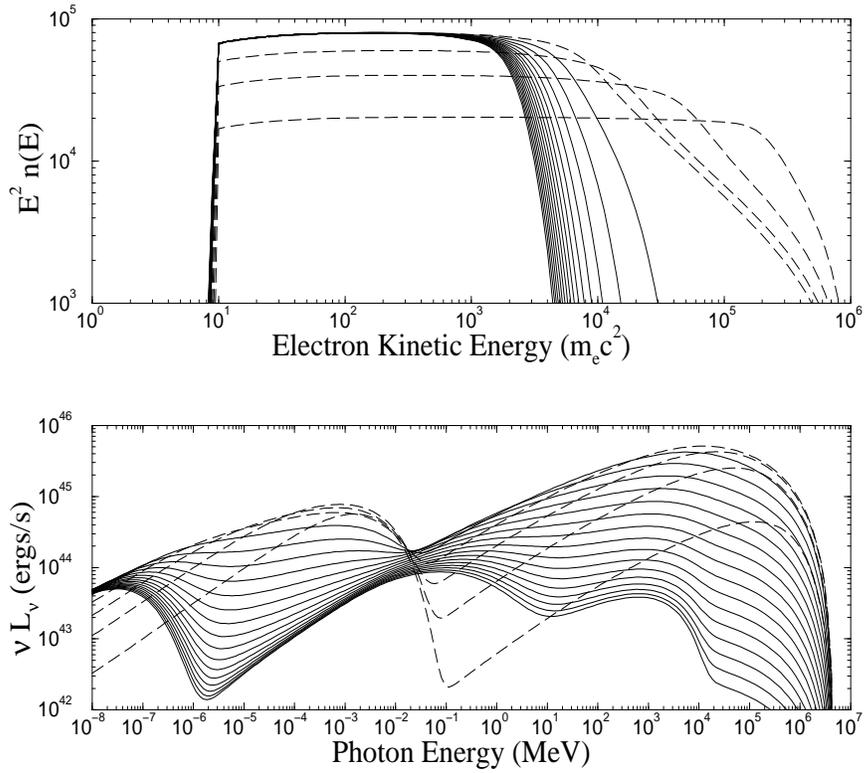,height=5in,width=4in,angle=-90}
\caption{Same as Figure \ref{fig:pp-e44}, except that
${\cal E} = 10^{49}$ ergs. The SSC component dominates over
the synchrotron component, and the buildup of synchrotron photon
energy-density is so quick that electron cooling is very efficient.
Towards the end of simulation ($\sim 10 t_{\rm dyn}$), multiple
Compton scattering features are evident.}
\label{fig:pp-e49}
\end{figure}

We now study in detail the full time evolution of three injection
cases: ${\cal E} = 10^{44}, 5\times 10^{47},$ and $10^{49}$ ergs.
They are shown in 
Figures \ref{fig:pp-e44}, \ref{fig:pp-5e47}, and \ref{fig:pp-e49},
where the time-evolution of 
electron distributions and photon spectra are presented.
Figures \ref{fig:ltcv-e44}, \ref{fig:ltcv-5e47}, and 
\ref{fig:ltcv-e49} show the corresponding light curves of different
photon energies for the above three injection cases. 
To qualitatively understand the light curves, 
we refer to equation (\ref{eq:dndt}) again. The peak
of the light curves is reached when the production and escape
are balanced, which depends on whether particle distribution
has softened enough. Furthermore, once the production at certain 
photon energy has
stopped, the pure escape process will produce an exponential decay.
Since the photon escape timescale is $t_{\rm dyn} = R/c$, 
one could get an 
estimate of the size by fitting the decline portion of the
light curves. This can be done using Figures \ref{fig:ltcv-e44}, 
\ref{fig:ltcv-5e47}, and \ref{fig:ltcv-e49} where fluxes have been
plotted in logarithm. The straight 
lines give a clear representation of photon escape, especially when it 
shows up in several energy bands, thus this might be a useful
method in analyzing the real blazar data.

\subsubsection{Spectral Variations}

The flux changes depicted in the above subsection are
accompanied by large spectral variations too as shown
in Figures \ref{fig:pp-e44}, \ref{fig:pp-5e47}, and \ref{fig:pp-e49}. 
These curves contain a wealth
of information, some of which are rather parameter dependent.
Nevertheless, we can draw some general conclusions:

(1) Since the cooling time at $\gamma_{\rm max}$ is the shortest
timescale in our system, the electron distribution at high energy
end is always substantially softened.
Furthermore, the production of photons $>$ tens of GeV is always
in the KN regime using SSC model. Both effects make the TeV spectrum very
soft, with a large curvature. [Additional effects such
as intrinsic absorption at the source or intergalactic 
absorption by infrared background can cause further curvature
(e.g., Coppi \& Aharonian 1999).]
This curvature is not consistent with the TeV spectra we have seen 
from Mrk 421 (\cite{ketal99}), but it is perhaps consistent with the
observations of Mrk 501 where a curvature in the Compton component is
clearly seen (\cite{cat99}). 

\begin{figure}
\epsfig{file=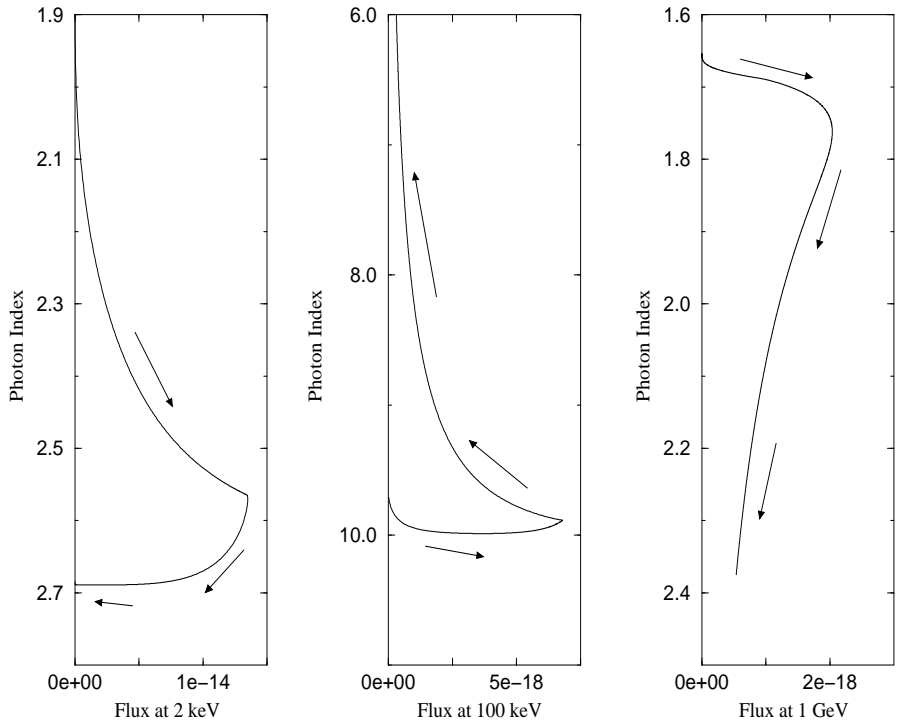,height=5in,width=4in,angle=-90}
\caption{The time evolution of the correlation between 
the photon index and the flux (ergs cm$^{-2}$ s$^{-1}$) at 2 keV, 100 keV, 
and 1 GeV, respectively.
Arrows indicate the direction of the time evolution.
The total injected energy is ${\cal E}=10^{44}$ ergs.
A large spectral evolution is seen at 100 keV, where
synchrotron and SSC components mix.
Spectral evolution at keV and GeV bands are relatively
moderate and clockwise.
}
\label{fig:flxinx-e44}
\end{figure}

\begin{figure}
\epsfig{file=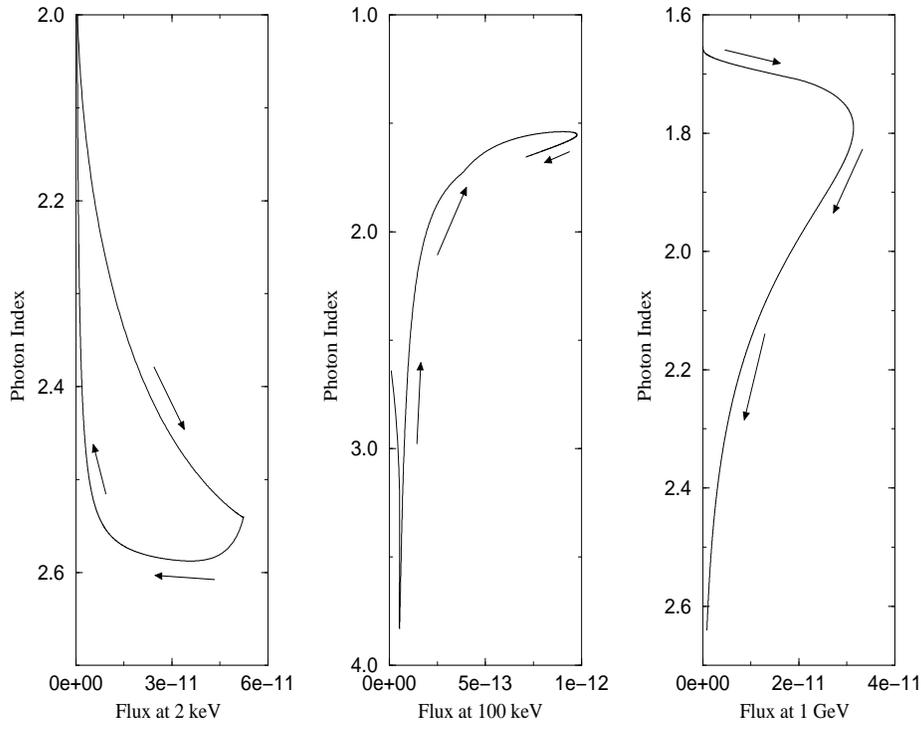,height=5in,width=4in,angle=-90}
\caption{Same as Figure \ref{fig:flxinx-e44}
except that ${\cal E} = 5\times 10^{47}$ ergs.
The evolution is qualitatively the same as in Figure \ref{fig:flxinx-e44}.
}
\label{fig:flxinx-5e47}
\end{figure}

\begin{figure}
\epsfig{file=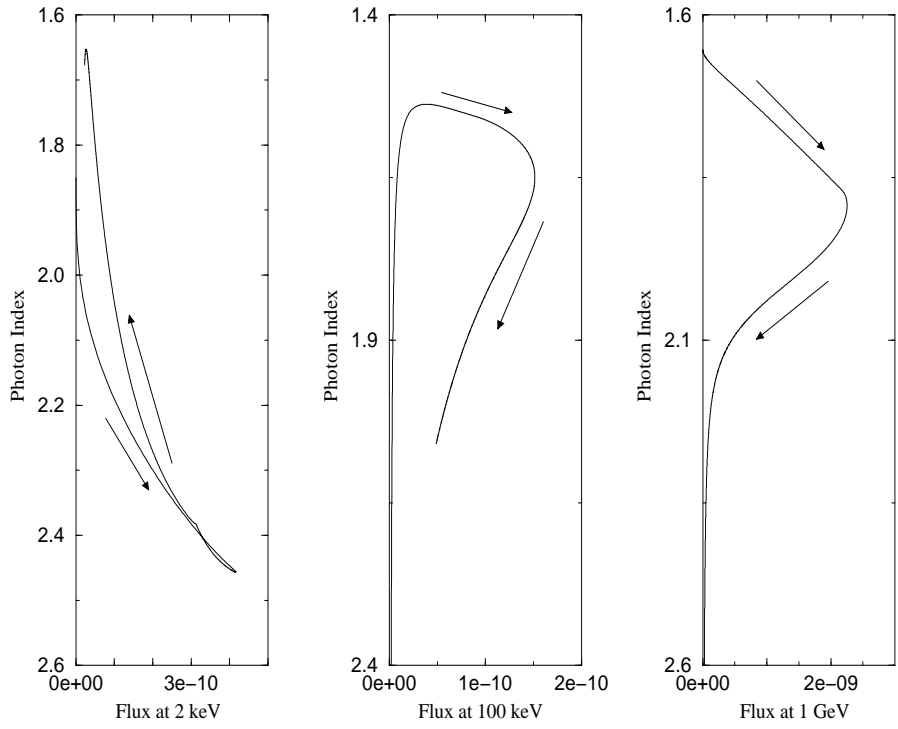,height=5in,width=4in,angle=-90}
\caption{Same as Figure \ref{fig:flxinx-e44}
except that ${\cal E} = 10^{49}$ ergs.
Strong evolution at 2 keV is evident, mostly
due to the efficient cooling of electrons by
SSC process.
}
\label{fig:flxinx-e49}
\end{figure}

(2) The ``hysteresis'' in the relation of
photon energy flux and spectral index was first pointed
out by Takahashi et al. (1996) at the keV band.
In Figures \ref{fig:flxinx-e44}, \ref{fig:flxinx-5e47}, and
\ref{fig:flxinx-e49}, we show 
the evolution of photon index as a function of energy flux 
in the observer's frame.
Clockwise rotation is always seen at 1 GeV, regardless
of whether synchrotron or SSC cooling dominates.
Clockwise rotation is found at 2 keV for
the case where synchrotron losses dominate the electron
cooling (${\cal E} = 10^{44}$ ergs). 
We find that this clockwise rotation at 2 keV is true for
all cases with injection energy less than $10^{48}$ ergs
(with the above injection form). 
This is mostly related to the fact that if we can associate
2 keV with the synchrotron peak, the spectrum always softens when its
flux is decreasing because the electrons that can produce 2 keV photons
have diminished. When the injection energy is large, the
SSC loss is dominant (${\cal E} = 10^{49}$ ergs),
the hysteresis diagrams rotate in the opposite sense at 2 keV.
The hardening at later time is actually due to the fact that
2 keV flux is now from the first generation of SSC, not synchrotron
anymore. Different behaviors of the hysteresis, however, are found at 
100 keV: counter-clockwise for ${\cal E} = 10^{44}$ ergs
and clockwise for ${\cal E} = 10^{49}$ ergs.
Given these large variations and their sensitivity
of various parameters, it is difficult to use these 
``hysteresis'' diagrams to draw firm conclusions.

(3) If observations show a synchrotron peak in 1 -- 10 keV band,
one should be very careful with fitting the spectrum in the 100 keV
energy band (such as OSSE and BeppoSAX), 
since it is right in the region where synchrotron and SSC meet. 
There is a very large and rapid spectral evolution during the flare
(e.g., Figure \ref{fig:flxinx-5e47}). 

(4) In the rising part of the $\nu F_\nu$ spectra, such as
1 -- 10 eV and MeV -- GeV, the spectral index variation is much
slower than the keV and TeV energy bands even though their fluxes vary 
by a large factor (e.g., Figures 
\ref{fig:flxinx-e44}, \ref{fig:flxinx-5e47}, and
\ref{fig:flxinx-e49}).

(5) To further quantify the spectral evolution, we
plot the peak energies of synchrotron and SSC components in 
$\nu F_\nu$ as a function of time in Figure \ref{fig:epkt}
for the case with ${\cal E} = 5\times 10^{47}$ ergs. An important
part is the early softening stage, where the synchrotron peak energy
decreases as $\gamma^2$ whereas the SSC peak energy goes down first
as $\gamma$ because the scattering is in the KN regime. 
This effect might be observable with the current keV and TeV
observational capabilities. Such a ``correlated'' evolution
in peak energies might provide a definitive test of SSC. 

(6) As shown in Figure \ref{fig:6curves}, when the SSC cooling
becomes comparable to or dominant over the synchrotron cooling,
the synchrotron peak becomes broader than those dominated
by synchrotron cooling only.

Some of the above conclusions might be testable using the
current data collected on blazars, and some might require 
much higher quality data.

\subsubsection{Time-Dependent Injection}

\begin{figure}
\epsfig{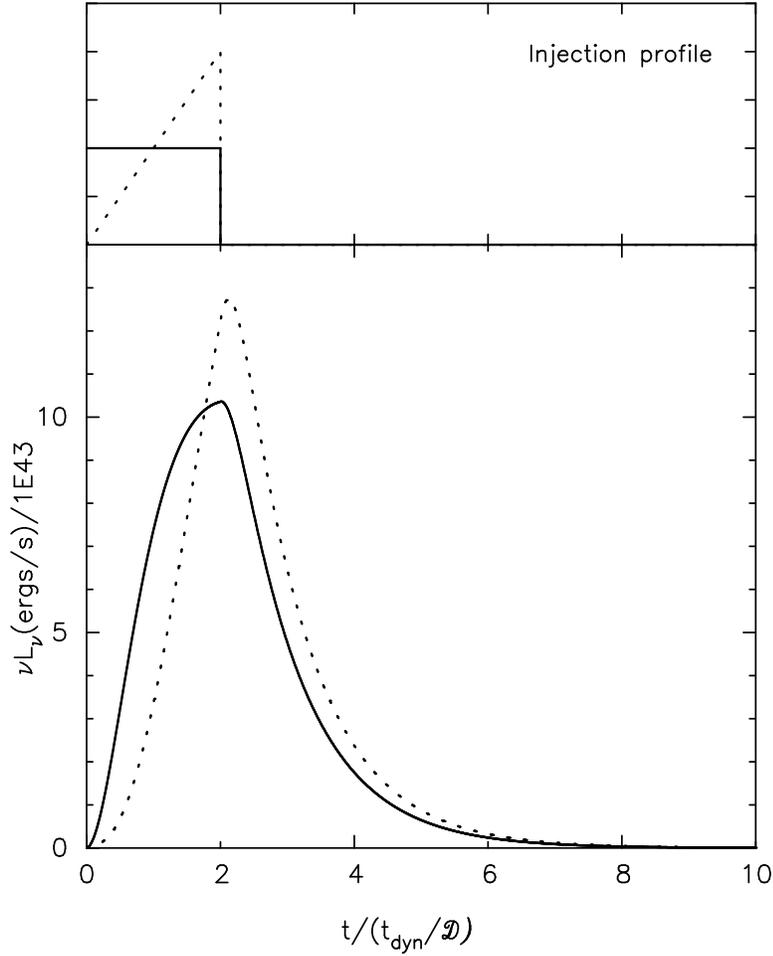}
\caption{
Light curves of observed spectra at 1 keV when power-law electrons
are injected.
Solid curve is calculated when
$Q(\gamma) = Q_0 \gamma^{-2} $ for
$0 \leq t \leq 2 t_{\rm dyn}$, and $Q(\gamma) = 0$ for
$t > 2 t_{\rm dyn}$.
Dotted curve is obtained 
when $Q(\gamma) = Q_0 \gamma^{-2} t/t_{\rm dyn}$ for
$0 \leq t \leq 2 t_{\rm dyn}$, and $Q(\gamma) = 0$ for
$t > 2 t_{\rm dyn}$.
The upper panel shows the time profile of 
the injection $Q$ in arbitrary units.
}
 \label{fig:light-triangle-box}
\end{figure}

In this subsection we show how different injection profiles
change the light curves in a single injection event.
Different from the previous subsection where a constant
injection is used (box-shaped injection), 
we calculate another case with 
a linearly increasing injection rate (triangle-shaped injection),
i.e., $Q(\gamma) =  Q_0 \gamma^{-2} t/t_{\rm dyn}$ for
$0 \leq t \leq 2 t_{\rm dyn}$, and $Q(\gamma) = 0$ for
$t > 2 t_{\rm dyn}$. 
The time evolution is followed until $t = 10 t_{\rm dyn}$
and electrons are allowed to escape with 
$t_{e, {\rm esc}} = 5 t_{\rm dyn}$.
We use the same number of particles and same amount of
total injected energy in both box- and triangle-shaped injections;
the injected energy is $5 \times 10^{47}$ ergs per $2 t_{\rm dyn}$
in the blob frame.
In Figure \ref{fig:light-triangle-box}, we compare 
their light curves at 1 keV.
The light curve for the box-shaped injection is asymmetric,
because more electrons are injected at an early stage
than in the triangle injection.
On the other hand, the light curve from triangle injection
shape is almost symmetric.
In both cases, light curves decay exponentially
after the end of the injection.

\subsubsection{Multiple Injections}
\label{sec:multi}

We now move to study other injection profiles, 
which are done by artificially turning 
the electron source term on and off. 
This is admittedly quite artificial.
The purpose is to understand whether there are any generic 
features associated with these multiple injections, which might aid us on 
modeling the multiple flares often observed from blazars.
In all the following runs, we allow electrons to escape with 
$t_{e, {\rm esc}} = 5 t_{\rm dyn}$.
Other parameters are the same as the single injection case.

\begin{figure}
\epsfig{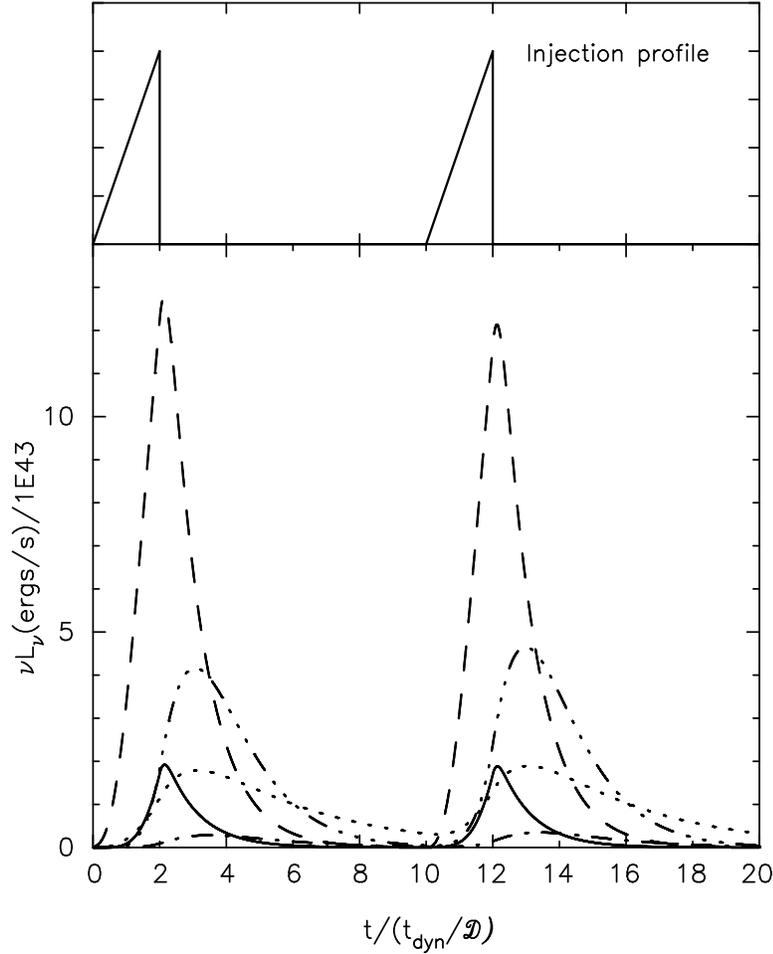}
\caption{Expected light curves at different wavelengths (lower panel)
when power-law electrons are injected according to the
profile in the upper panel. The time is measured in the observer's
frame. In the comoving frame, two injections are separated by 
$8 t_{\rm dyn}$, longer than the electron escape timescale, 
which is chosen as $5 t_{\rm dyn}$.  The same amount of energy,
$5 \times 10^{47}$ ergs, is injected in each flare.
The dotted, dashed, dash-dotted, dash-dot-dot-dotted, and
solid represent fluxes at 1 eV, 1 keV, 1 MeV, 1 GeV, and 1 TeV,
respectively. The two flares can be regarded as a simple sequence of
two unrelated injections.
}
 \label{fig:multi-light-long}
\end{figure}

\begin{figure}
\epsfig{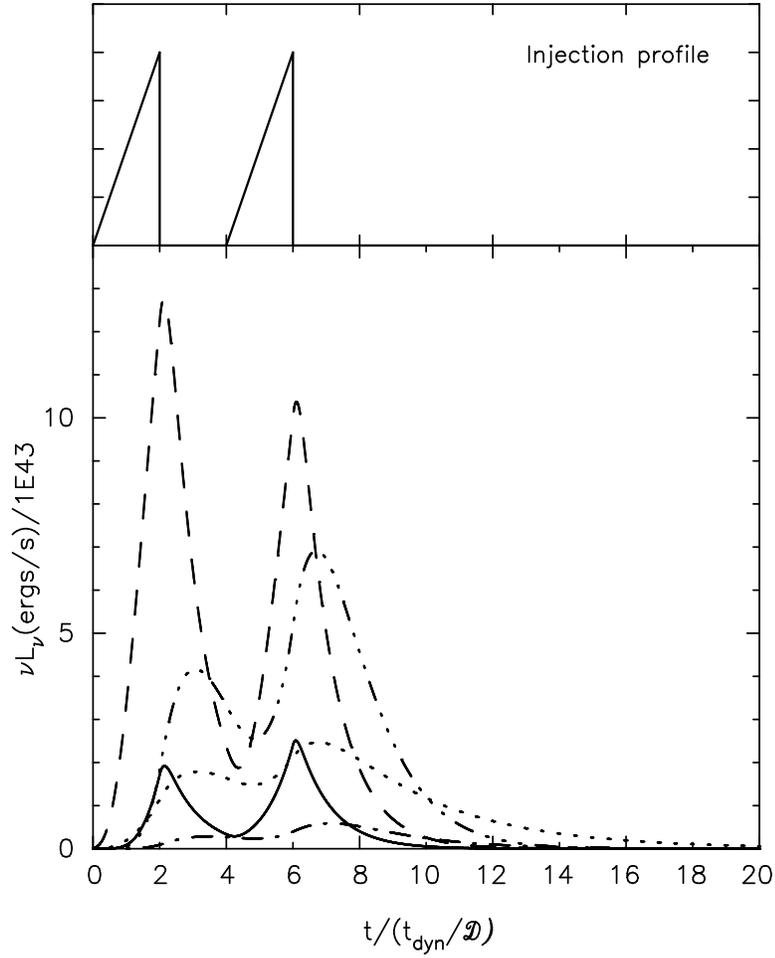}
\caption{Same as Figure \ref{fig:multi-light-long},
except that the separation of two flares is shorter ($2 t_{\rm dyn}$)
than the electron escape timescale. The second flare is now strongly
affected by the residual effects from the first electron injection. 
}
 \label{fig:multi-light-short}
\end{figure}

Successive flares can be produced by repeated injections
of nonthermal electrons. 
As an example of this picture, 
we present light curves for two repeated injection cases
of nonthermal electrons.
The top panels of Figures \ref{fig:multi-light-long} and
\ref{fig:multi-light-short} show the injection profiles,
which consist of two triangle injections separated by a
long ($8 t_{\rm dyn}$) and a short ($2 t_{\rm dyn}$) intervals
(in the comoving frame), respectively.
In both cases $5 \times 10^{47}$ ergs are injected in each ``flare''
with a triangle-shaped time profile.
The lower panels of Figures \ref{fig:multi-light-long}
and \ref{fig:multi-light-short} show the expected
light curves calculated by our code.
The shape of the light curves from each injection is very
similar to that in Figure \ref{fig:light-triangle-box}
where a single injection is involved (i.e., quasi-symmetric).
The peak fluxes of light curves in multiple injections are, however,
affected by the separation time of two flares.
When the separation interval is longer than the electron escape time
($5 t_{\rm dyn}$), the light curves can almost be regarded as a simple
sequence of two separate single injections. But when the
separation interval is shorter than the electron escape time,
multiwavelength light curves become rather complicated.
The main physical reason behind this complication is the
dynamic accumulation of both photons and relativistic 
electrons. First of all, 1 eV and 1 keV emissions are from synchrotron
and others are from SSC. All the SSC emissions have second peak
higher than the first one, this is due to the increase both
in soft photon energy density and in the number of electrons which
are not completely cooled yet when the second injection occurs.
This accumulation of electrons also accounts for the increase in
1 eV synchrotron flux. The 1 keV emissions, however, show a lower flux
in the second peak, this is because the relativistic
electrons from the second injection are subject to a much
stronger cooling due to the enhanced photon energy density from
the first injection. In addition to the flux differences,
there are obvious delays in reaching the peak fluxes
for different wavelengths with respect to the synchrotron and SSC
peaks, though 1 keV and 1 TeV fluxes track each other rather well.

As demonstrated in these figures, the slow response and relatively
small amplitude variations at the photon energies other than the
synchrotron and SSC peaks argue against the usual belief of closely 
correlated variations in multiwavelength observations. 
Only the emissions from the tail of
the electron energy distribution can be reliably used as diagnostics
for separate injections. Still, extra caution is obviously needed 
when relating the energy contained in nonthermal particles versus
the observed fluxes.

\section{SUMMARY AND DISCUSSIONS}
\label{sec:sum}

Using a homogeneous synchrotron self-Compton model, 
we have calculated the time evolution of emission spectra
and electron energy distributions when nonthermal electrons
are uniformly injected into a relativistically moving plasma blob with
constant velocity. 
We have found that:

(1) When the luminosities of the synchrotron and SSC peaks
are comparable, the electron cooling by inverse Compton scattering
is not negligible and the system is inherently nonlinear
and dynamic. One has to solve the time-dependent,
coupled electron and photon kinetic-equations self-consistently.
Furthermore, since observations
are most sensitive to the peak fluxes of synchrotron and 
SSC components, accurate treatments of synchrotron emissivity
due to the end point effects and inverse
Compton scattering in the KN regime are quite essential.
 
(2) When the cooling time at the maximum particle energy is
longer than the  injection
timescale ($\ge R/c$), the light curve of emissions
corresponding to the tail of the electron distribution can
have short, large amplitude variations but emissions
at other wavelengths show considerably longer and 
smaller amplitude changes. Additionally, spectral
evolution is also rather slow. All these features are 
simply caused by the long cooling time of electrons.

(3) When cooling time at the maximum particle energy is
shorter than the dynamic timescale, strong spectral
evolutions are observed for both synchrotron and SSC 
components and short duration flares are obtained in
most energy bands.

(4) Generally, the prompt TeV spectrum is curved due to the
KN effect and the fact that TeV-production electrons
are usually in a cooled distribution. This consideration
does not take into account the possible infrared background
attenuation of the TeVs, which might cause further curvature
in the TeV spectrum. On the other hand, most current TeV
observations require an accumulation time probably much
longer than the dynamic timescale of the blob, so that it
might still be possible to obtain a quasi power-law TeV
spectrum by averaging over an evolving spectrum. Further
study is needed to address this issue.

\begin{figure}
\epsfig{file=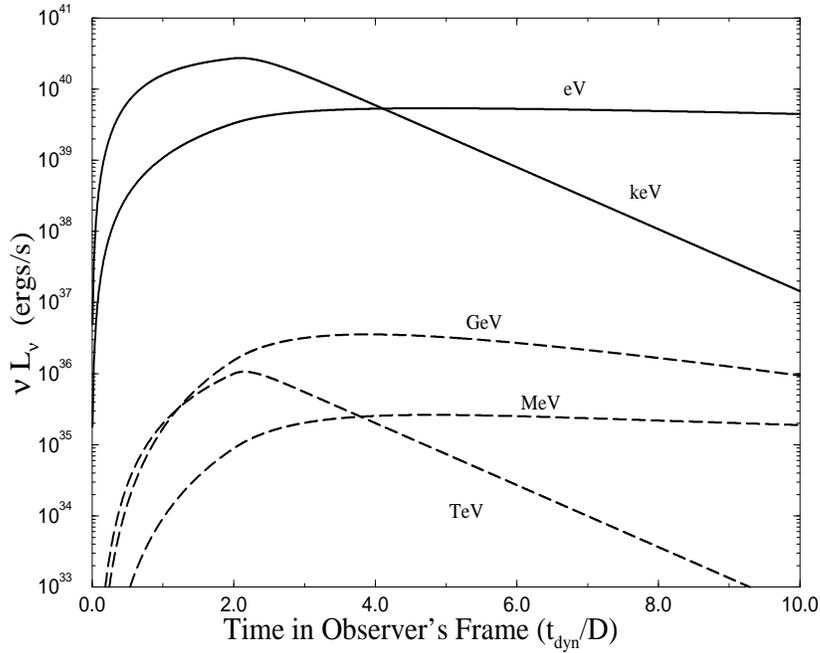,height=5in,width=4in,angle=-90}
\caption{Multiwavelength light curves in observer's frame
(${\cal D}$ is assumed to be 10) with ${\cal E} = 10^{44}$ ergs
(cf. Figure \ref{fig:pp-e44}).
Solid and dashed curves are for synchrotron and SSC components,
respectively. The exponential decay depicted by the keV and TeV
fluxes allows a direct estimate of the size of the emission cloud.
}
\label{fig:ltcv-e44}
\end{figure}

\begin{figure}
\epsfig{file=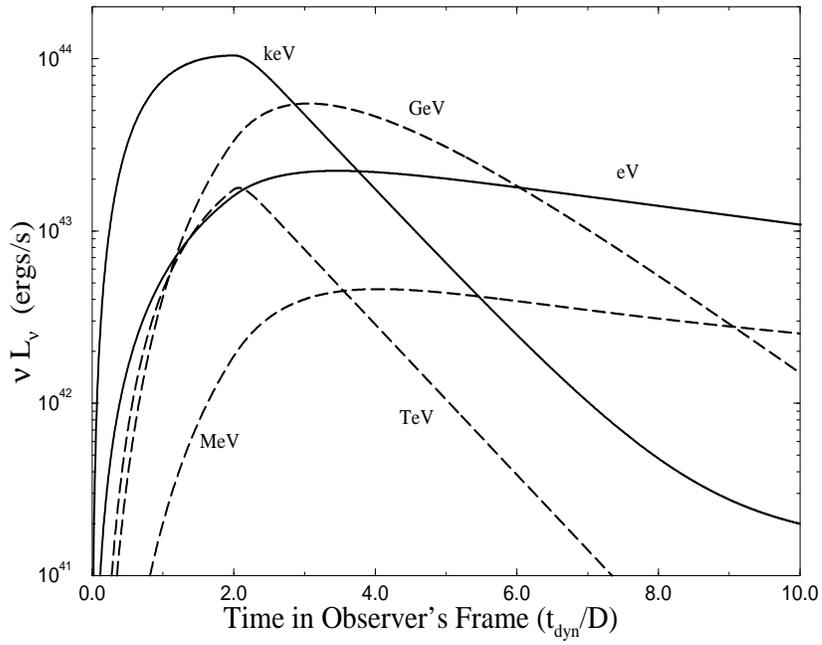,height=5in,width=4in,angle=-90}
\caption{Same as Figure \ref{fig:ltcv-e44}
except that ${\cal E} = 5\times 10^{47}$ ergs. 
(Also cf. Figure \ref{fig:pp-5e47})}
\label{fig:ltcv-5e47}
\end{figure}

\begin{figure}
\epsfig{file=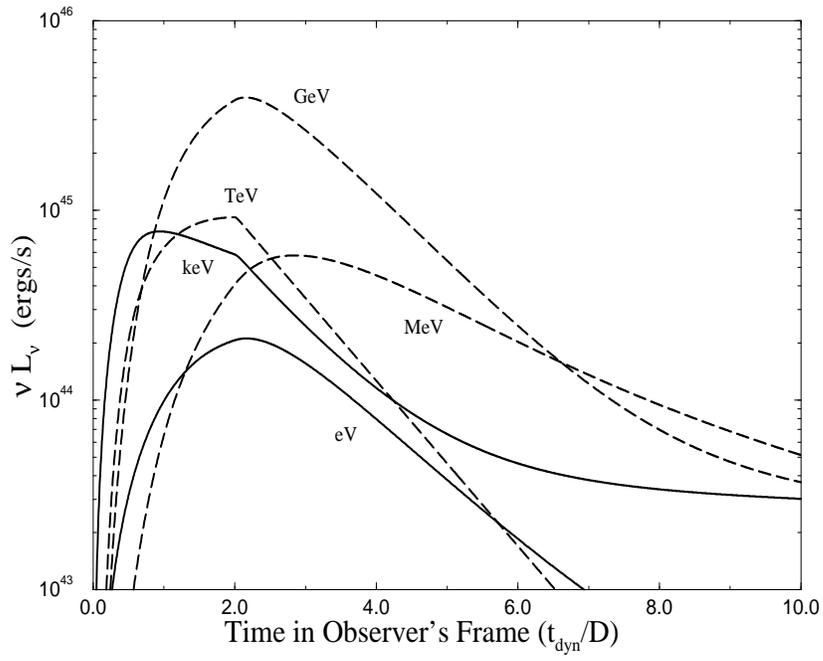,height=5in,width=4in,angle=-90}
\caption{Same as Figure \ref{fig:ltcv-e44}
except that ${\cal E} = 10^{49}$ ergs. 
(Also cf. Figure \ref{fig:pp-e49})}
\label{fig:ltcv-e49}
\end{figure}

(5) We recommend plotting the light curves in a fashion
that is logarithmic flux versus linear time. The goal
is to look for exponential decays at specific energy
bands, which might give direct measurements of the size
of the system, as indicated in Figures
\ref{fig:ltcv-e44} -- \ref{fig:ltcv-e49}.

(6) One has to be cautious about the common belief that 
light curves in different energy bands should track each other.
The electrons responsible for producing specific energy photons
might have quite different lifetimes, especially when
multiple and closely spaced injections are involved. 
This complication also applies to the leading/lagging
analysis for different photon energy bands.

\begin{figure}
\epsfig{file=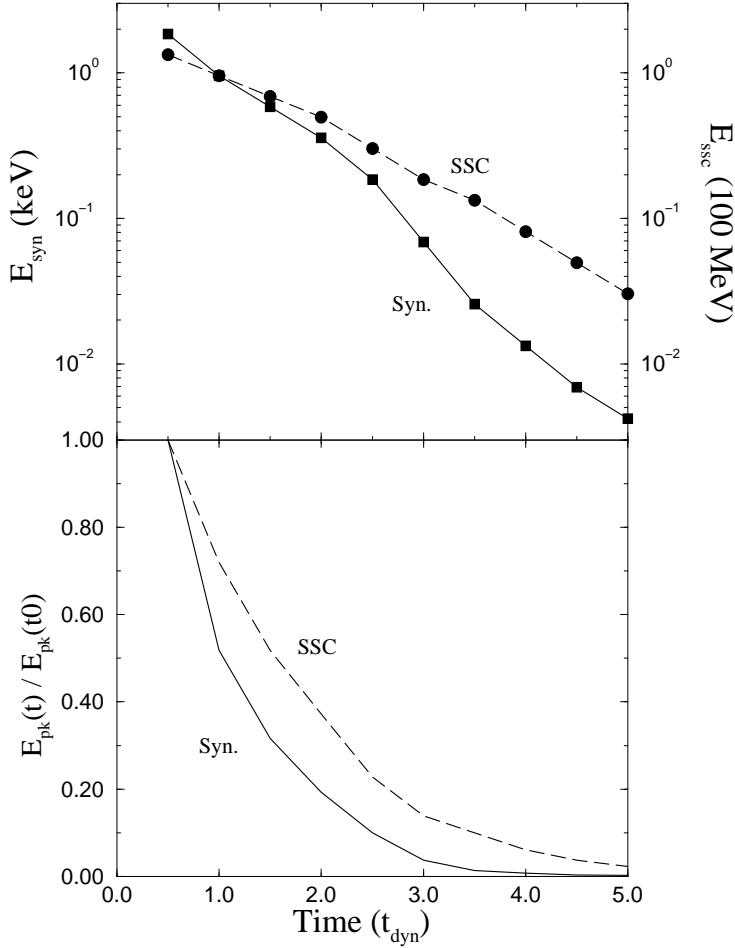,height=5in,width=4in,angle=0}
\caption{Evolution of peak photon energies of both synchrotron and SSC
components. The total injected energy is ${\cal E}=5\times 10^{47}$ ergs
(cf. Figure \ref{fig:pp-5e47}). 
The upper panel shows their actual values and the lower panel
shows the normalized values after dividing each peak energy by the
value at $t = 0.5 t_{\rm dyn}$ for synchrotron and SSC components,
respectively. The synchrotron peak energy ($\propto \gamma^2$)
decreases faster than SSC peak energy ($\propto \gamma$) {\em initially}, 
because SSC process is still in the Klein-Nishina regime. 
}
\label{fig:epkt}
\end{figure}

(7) When high time-resolution spectroscopy is available
both in keV and TeV bands, one should be able to prove
whether TeV production is via SSC process by comparing
the rates of spectral softening as done in Figure \ref{fig:epkt}.

The primary purpose of this paper is to investigate 
the radiative signatures in a purely cooling and dynamic system,
thus providing a bridge between observations and the detailed
but largely unknown physics of particle energization processes.
Since we did not address the particle acceleration problem here,
in this sense, some of the conclusions drawn above are certainly
subject to revisions as our understanding of energy flow in
AGNs improves.

In conclusion, we have found that solving time-dependent, coupled 
electron and photon kinetic-equations provides an easy
and efficient way of comparing multiwavelength, time-dependent
observations with some simplified SSC models. It has the
advantage of naturally combining the spectral and temporal
evolutions in a dynamic system, which is very useful when more
and more high quality data become available.

\acknowledgements

We thank C. Dermer for useful discussions and 
the anonymous referee for the helpful comments.
H.L. gratefully acknowledges the support of an Oppenheimer Fellowship, 
and his research is supported by the Department of Energy, 
under contract W-7405-ENG-36.
M.K. thanks F. Takahara for stimulating discussions and
his research was partially supported by Scientific Research Grants
(09223219, 10117215) from the Ministry of Education, 
Science, Sports and Culture of Japan.

\clearpage

\begin{thebibliography}{}

\bibitem[Blandford \& Rees 1978]{br78}
Blandford, R.D., \& Rees, M.J. 1978, in Pittsburgh Conf. on BL Lac
Objects, ed. A.M. Wolfe (Pittsburgh: Univ. Pittsburgh Press), 328

\bibitem[Bloom \& Marscher (1996)]{bm96}
Bloom, S. D., \& Marscher, A. P. 1996, \apj, 461, 657

\bibitem[]{}
B$\ddot{o}$ttcher, M., Mause, H., \& Schlickeiser, R. 1997,
A\&A, 324, 395

\bibitem[Buckley et al. (1996)]{buck96}
  Buckley, J. H. et al. 1996, \apjl, 472, L9
\bibitem[Catanese et al. (1997)]{cat97}
  Catanese, M. et al. 1997, \apjl, 487, L143

\bibitem[Coppi \& Aharonian 1999]{ca99}
  Coppi, P. S., \& Aharonian, F. A. 1999, \apjl, 521, L33

\bibitem[Coppi \& Blandford 1990]{cb90}
  Coppi, P. S., \& Blandford, R. D. 1990, \mnras, 245, 453

\bibitem[Crusius \& Schlickeiser (1986)]{cs86}
  Crusius, A., \& Schlickeiser, R. 1986, A\&A, 164, L16

\bibitem[Dermer (1998)]{dermer98}
  Dermer, C. D. 1998, \apjl, 501, L157

\bibitem[Dermer, Sturner, \& Schilickeiser (1997)]{dss97}
  Dermer, C. D., Sturner, S. J., \& Schlickeriser, R. 1997,
   \apjs, 109, 103

\bibitem[Djannati-Atai et al. 1999]{cat99}
  Djannati-Atai, A. et al. 1999, A\&A, submitted

\bibitem[Georganopoulos, \& Marcher (1998)]{gm98}
  Georganopoulos, M., \& Marcher, A. P.1998, \apjl, 501, L11

\bibitem[Ghisellini et al. (1988)]{ghi88}
  Ghisellini, G., Guilbert, P. W., \& Svensson, R. 1988
  \apjl, 334, L5

\bibitem[ghisellini \& Madau (1996)]{gm96}
  Ghisellini, G., \& Madau, P. 1996, \mnras, 280, 67

\bibitem[Ghisellini et al. (1998)]{gcfmc98}
  Ghisellini, G., Celotti, A., Fossati, G.,
  Maraschi, L., \& Comastri, A. 1998, \mnras, 301, 451

\bibitem[Inoue \& Takahara 1996]{it96}
  Inoue, S., \& Takahara, F. 1996, \apj, 463, 555

\bibitem[Jones (1968)]{jones68}
  Jones, F. C. 1968, Physical Review, 167, 1159

\bibitem[Kataoka et al. (1999)]{kat99}
  Kataoka, J. et al. 1999, \apj, 514, 318

\bibitem[Kirk et al. (1998)]{kirketal98}
  Kirk, J. G., Rieger, F. M., \& Mastichiadis, A. 1998, \aap, 333, 452

\bibitem[Krennrich et al. 1999]{ketal99}
  Krennrich, F. et al. 1999, \apj, 511, 149

\bibitem[Kusunose, Coppi, \& Li 1999]{kcl99}
  Kusunose, M., Coppi, P.S., \& Li, H. 1999, unpublished,
available upon request

\bibitem[Macomb et al. (1995)]{macomb95}
  Macomb, D. J., et al. 1995, \apjl, 449, L99

\bibitem[Macomb et al. (1996)]{macomb96}
  Macomb, D. J., et al. 1996, \apjl, 459, L111 (erratum)

\bibitem[Mastichiadis \& Kirk (1997)]{mk97}
  Mastichiadis, A., \& Kirk, J. G. 1997, \aap, 320,19

\bibitem[Pian et al. (1998)]{pian98}
  Pian, E. et al. 1998, \apjl, 492, L17

\bibitem[Robinson \& Melrose (1984)]{rm84}
  Robinson, P. A., \& Melrose, D. B. 1984, Australian J. Physics, 37, 675

\bibitem[Romanova \& Lovelace 1997]{rl97}
  Romanova, M.M. \& Lovelace, R.V.E. 1997, \apj, 475, 97

\bibitem[Sikora et al. (1997)]{smmp97}
  Sikora, M., Madejski, G., Moderski, R., Poutanen, J. 1997,
  \apj, 484, 108

\bibitem[Takahashi et al. (1996)]{taka96}
  Takahashi, T., Tashiro, M., Madejski, G., Kubo, H.,
  Kamae, T., Kataoka, J., Kii, T., Makino, F.,
  Makishima, K., \& Yamasaki, N. 1996, \apjl, 470, L89

\bibitem[Vermeulen \& Cohen 1994]{vc94}
  Vermeulen, R.C. \& Cohen, M.H. 1994, \apj, 430, 467

\bibitem[von Montigny et al. 1995]{vm95}
  von Montigny, C. et al. 1995, \apj, 440, 525

\end{thebibliography}
\end{document}